\documentclass[10pt,a4paper]{article}
\usepackage[latin1]{inputenc}
\usepackage{epsf}
\usepackage[body={6.5 in,9.0in},left=1.0 in,top=1.0 in]{geometry}
\usepackage{amsmath}
\usepackage{amsfonts}
\usepackage{amssymb}
\usepackage{graphicx}
\usepackage{multicol}
\usepackage{makeidx}
\begin{document}
\author{Subhankar Roy\footnote{Email: meetsubhankar @ gmail.com} and N. Nimai Singh\footnote{E-mail: nimai03 @ yahoo.com}\\ \\
Department of Physics, Gauhati University, Guwahati-781014, India }
\title{Bi-Large neutrino mixing with charged lepton correction.}
\maketitle
\abstract
The usual Bi-Maximal (BM) neutrino mixing faces an inherent problem in lowering the solar angle below $\tan^{2}\theta_{12}=0.50$ when charged lepton correction is taken. This minimum  $\theta_{12}$ is achievable only if $CP$ violation is absent. We start with a new model which incorporates a new idea of mixing devoloped recently, called Bi-Large (BL) mixing, similar to BM mixing except that the former chooses rather $\theta_{13}$ as Cabibbo angle ($\theta_{c}$) than zero. We apply this mixing in the neutrino sector followed by a charged lepton correction with the CKM type matrix $U_{l}$. The model marks a  prediction on $\theta_{23}$ to lie within the first octant. The $CP$ violating phase $\delta_{CP}$ dictates the prediction of all the three mixing angles. A proper choice of $\delta_{CP}$, leads to the predictions of all the three mixing angles including $\theta_{12}$, to align very precisely with the experimental bestfit. This close agreement thus hoists Bi-Large mixing as an important and promising mixing scheme, in contrast to BM or TBM mixing as a first approximation. A formal derivation of BL mixing from discrete symmetry will be an important investigation in neutrino physics.
\\
\\
PACS numbers: 14.60.Pq, 12.15.Ff.\\
\\
Keywords: Bi-Large mixing, Bi-maximal mixing, Tri-bimaximal mixing, neutrino masses, Cabibbo angle, CKM matrix, PMNS Matrix .
\section{Introduction}
The recent experimental data from Double Chooz [1], Daya Bay [2], RENO [3], T2K [4] and MINOS [5] collaborations, indicate not only a nonzero reactor angle ($\theta_{13}$) but also with its magnitude of the order of Cabibbo angle ($\theta_{c}$). Tri-Bimaximal (TBM) mixing [6] and Bimaximal (BM) mixing [7, 8, 9] are two popular mixing patterns which predict $\sin\theta_{13}=0$. TBM mixing has a strong theoretical support because of its relation with $A_{4}$ [10 - 14], one of the candidates of discrete flavour symmetry groups. From theoretical point of view, small deviation of the order of square of $\lambda_{c}$ ( where $\lambda_{c} = \sin\theta_{c} \approx 0.22$ ) is expected. But a large correction of the order of $\lambda_{c}$ to $\sin\theta_{13}=0$, clearly interrogates the loyality of TBM mixing as a first approximation. This was pointed out in the literature [15]. The same argument holds good for BM mixing scheme also. In addition, at the Neutrino 2012 conference the MINOS collaboration hinted for a non-maximal $\theta_{23}$, which also goes against the TBM and BM predictions. From the analyses given in Ref. [16, 17], $\theta_{23}$ tilts towards a preference for $\theta_{23}< 45^{0}$.  

 A new idea of mixing scheme called Bi-Large (BL) mixing [15] has been proposed recently by Boucenna $et.al$, apart from the existing TBM and BM mixing schemes. They  considered $\sin\theta_{13}$ as the fundamental parameter ($\lambda$) and the idea behind this ansatz lies in the smallness of $\theta_{13}$, among the three mixing parameters. They expressed $\sin\theta_{12}$ and $\sin\theta_{23}$ as linear functions of  $\lambda$. Thus, 
\begin{equation}
\sin\theta_{13}=\lambda,\quad
\sin\theta_{12}=a\lambda, \quad
\sin\theta_{23}=s\lambda .
\end{equation} 
Here $a$, $s$ are free parameters and $a \simeq s$. The resulting parametrization neither terminates to TBM nor BM pattern as limiting cases, though maximal atmospheric angle can be obtained. When $\lambda \rightarrow 0$, the neutrinos are unmixed. From simple numerical analysis they have shown that strict BL mixing occurs when $\lambda \simeq \lambda_{c} \approx 0.22$ and under that condition we get $ a = s =3 $. 

We start with this strict BL ansatz [Eq.(1)] where the Cabibbo angle ($\lambda_{c}$), the most important parameter from CKM matrix generates the whole parametrization in the neutrino sector.We take,
\begin{eqnarray}
\sin\theta_{13}=\lambda_{c},\quad
\sin\theta_{12}=3\lambda_{c}, \quad
\sin\theta_{23}=3\lambda_{c} \quad .
\end{eqnarray}
Pending a formal derivation of the BL mixing from a discrete symmetry, we wish to explore its matrix form  from phenomenological ground.Following the  standard PDG scheme of parametrization, we arrive at the following strict BL mixing matrix ($U_{BL}$),
\begin{align}
U_{BL} &=\begin{pmatrix}
 \frac{3}{4}(1-\frac{\lambda_{c}^{2}}{2})& \frac{\sqrt{7}}{4}(1-\frac{\lambda_{c}^{2}}{2})  & \lambda_{c}  \\ 
 -\frac{3\sqrt{7}}{16}(1+\lambda_{c})& \frac{9}{16}(1-\frac{7}{9}\lambda_{c}) & \frac{\sqrt{7}}{4}(1-\frac{\lambda_{c}^{2}}{2})   \\ 
 \frac{7}{16}(1-\frac{9}{7}\lambda_{c})& -\frac{3\sqrt{7}}{16}(1+\lambda_{c}) & \frac{3}{4}(1-\frac{\lambda_{c}^{2}}{2})
\end{pmatrix}, \\
 \nonumber\\
\quad \quad &=\begin{pmatrix}
 0.7309& 0.6446 & 0.2257 \\ 
 -0.608& 0.4637 & 0.6446 \\ 
 0.3105& -0.608  & 0.7309
\end{pmatrix}. \nonumber
\end{align}
$U_{BL}$ satisfies unitarity condition. If we first approximate the neutrino mixing matrix $U_{\nu}$ to $U_{BL}$ and then in order to account for the required deviations, we consider the correction from charged lepton sector [18]. We try to find out the possible texture of charged lepton matrix $U_{l}$ ( must follow unitarity condition ), which may serve our purpose.
\section{The problems in Bi-maximal (BM) mixing}
The strict BL mixing [15] and BM mixing patterns have certain similarities. $\theta_{12}$ and $\theta_{23}$ are equal for both the cases. The former predicts them to be $41^{0}$  and the later takes them as maximal i.e., $45^{0}$. The significant difference lies in the fact that the former starts with $\theta_{13}=\theta{c}$, and later with $\theta_{13}=0^{0}$,
\begin{eqnarray}
U_{BM}=\begin{pmatrix}
\frac{1}{\sqrt{2}}&\frac{1}{\sqrt{2}} & 0  \\ 
 -\frac{1}{2}& \frac{1}{2} & \frac{1}{\sqrt{2}} \\ 
 \frac{1}{2} & -\frac{1}{2} &\frac{1}{\sqrt{2}} \end{pmatrix}.
\end{eqnarray}
In the reference [18], the authors put forward an viable technique to comply with the experimental data. This is summarised as follows. In fact they considered $U_{\nu}= U_{BM}$ and then performed a charged lepton correction by choosing the charged lepton matrix $U_{l}$ to be CKM type,
\begin{eqnarray}
U_{l}=\begin{pmatrix}
1-\frac{\lambda_{c}^{2}}{2} & \lambda_{c} e^{i\delta_{cp}} & 0 \\ 
 -\lambda_{c} e^{-i\delta_{cp}} & 1-\frac{\lambda_{c}^{2}}{2} & 0 \\ 
0 & 0 & 1
\end{pmatrix}. 
\end{eqnarray}
The possible inclusion of Dirac phase $\delta_{cp}$ in $1-2$ and $ 2-1$ positions of $U_{l}$ was first introduced by Fritzsch and Xing [19]. In eq (5), $U_{l}$ satisfies unitarity condition. The $U_{PMNS}=U_{l}^{\dagger} U_{\nu}$ becomes,
\begin{eqnarray}
U_{PMNS}=\begin{pmatrix}
\frac{1}{\sqrt{2}}(1-\frac{\lambda_{c}^{2}}{2})+\frac{\lambda_{c}}{2}e^{i\delta_{cp}} & \frac{1}{\sqrt{2}}(1-\frac{\lambda_{c}^{2}}{2})-\frac{\lambda_{c}}{2}e^{i\delta_{cp}} & -\frac{\lambda_{c}}{\sqrt{2}}e^{i\delta_{cp}} \\ 
\frac{1}{2}(\frac{\lambda_{c}^{2}}{2}-1)+\frac{\lambda_{c}}{\sqrt{2}}e^{-i\delta_{cp}}  &\frac{1}{2}(1-\frac{\lambda_{c}^{2}}{2})+\frac{\lambda_{c}}{\sqrt{2}}e^{-i\delta_{cp}}   & \frac{1}{\sqrt{2}}(1-\frac{\lambda_{c}^{2}}{2}) \\ 
  \frac{1}{2} & -\frac{1}{2} &\frac{1}{\sqrt{2}}
\end{pmatrix} .
\end{eqnarray}
From Eq.(6), using the following relations, 
\begin{equation}
\sin^{2}\theta_{13} = \vert U_{e3}\vert ^{2}, \quad \sin^{2}\theta_{12} = \frac{\vert U_{e2}\vert ^{2}}{1- \vert U_{e3}\vert ^{2}}, \quad \sin^{2}\theta_{23} = \frac{\vert U_{\mu 3}\vert ^{2}}{1- \vert U_{e3}\vert ^{2}},
\end{equation}
we obtain
\begin{eqnarray}
\sin^{2}\theta_{13}&=& \frac{\lambda_{c}^{2}}{2}\approx 0.0254, \\
\sin^{2}\theta_{12} &=&\frac{4-2 \lambda_{c}^{2}+2 \sqrt{2}\lambda_{c}(\lambda_{c}^{2}-2)\cos\delta_{cp}}{8(1-\frac{\lambda_{c}^{2}}{2})} ,\\
\sin^{2}\theta_{23} &=&\frac{1}{2}(1-\frac{\lambda_{c}^{2}}{2})\approx 0.488,\\
J^{BM}_{CP}& \approx &\frac{1}{4\sqrt{2}} \lambda_{c} \sin\delta_{cp}.
\end{eqnarray}
The prediction of $\theta_{13}$ matches with the best-fit value [20], while that for $\theta_{23}$ lies within $2\sigma$ [20]. The prediction of $\theta_{12}$ depends on $\delta_{cp}$. Now if we want $\sin^{2}\theta_{12}$ as 0.32 (best-fit) [20], from Eq.(9), we have $\cos\delta_{cp}=1.13$, which is absurd. The relation between $\sin^{2}\theta_{12}$ and $\delta_{cp}$ is illustrated in Fig.1. The minimum value of 0.3407 for $\sin^{2}\theta_{12}$ (i.e.,$\tan^{2}\theta_{12}\approx 0.52$ ) is obtained at the cost of $\cos\delta_{cp}=1$,  which in turn gives CP violation parameter Jarkslog invariant $J^{BM}_{CP}=0$. This is the discrepency of BM model where $\sin^{2}\theta_{12}$ can not be suppressed even though $J_{CP}$ is sacrificed.
\begin{figure}
\begin{center}
\includegraphics[scale=1]{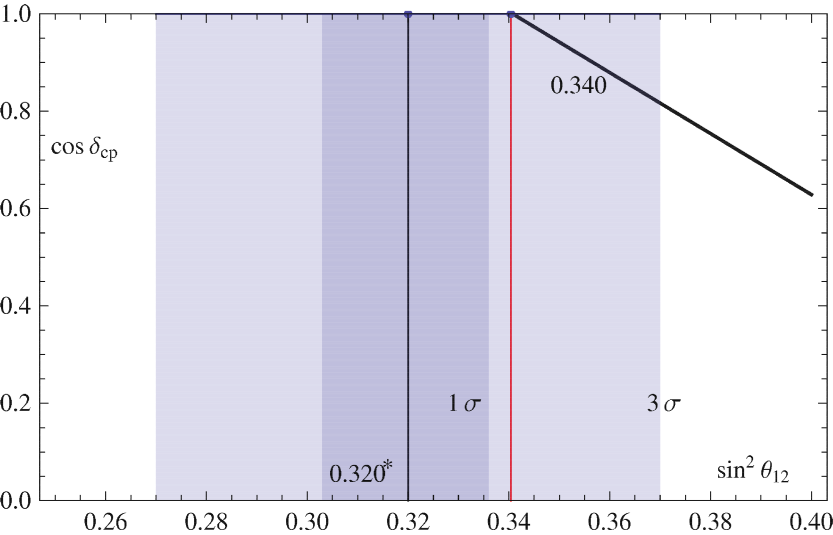} 
\caption{\footnotesize The dependence of $\cos\delta_{cp}$ on $\sin^{2}\theta_{12}$ for BM case with charged lepton correction. The prediction of  the solar angle can not be lowered to the present experimental best-fit through any possible way. The lowering of $\theta_{12}$ upto certain level is possible at the cost of $\delta_{cp}\rightarrow 0, 2\pi $.}
\end{center}
\end{figure}
\section{Strict Bi-Large mixing and Charged lepton contribution}
We now assume that neutrino mixing matrix $U_{\nu}$ follows strict BL mixing [Eq.(2), Eq.(3)] and take $U_{\nu} = U_{BL}$. We assume the charged lepton mixing matrix to be CKM type. Motivated by the similarities among the two mixing schemes and the partial success, we try with the same CKM type $U_{l}$ employed for BM case (Eq.(5)) [18]and generate $U_{PMNS} = U_{l}^{\dagger} U_{BL}$. 
\begin{eqnarray}
U_{PMNS}=\begin{pmatrix}
U_{e1} & U_{e2}  & U_{e3} \\ 
U_{\mu 1} & U_{\mu 2} & U_{\mu 3} \\ 
U_{\tau 1} & U_{\tau 2} & U_{\tau 3}.
\end{pmatrix},
\end{eqnarray}
where,\begin{eqnarray}
U_{e1}&=& \frac{3}{16}\lbrace(\lambda_{c}^{2}-2)^{2}+\sqrt{7}\lambda_{c}(1+\lambda_{c})e^{i \delta_{cp}}\rbrace, \nonumber\\
U_{e2}&=& \frac{1}{16}\lbrace \sqrt{7}(\lambda_{c}^{2}-2)^{2}+\lambda_{c}(7\lambda_{c}-9)e^{i \delta_{cp}}\rbrace , \nonumber\\
U_{e3}&=& \frac{1}{8} \lambda_{c}(\lambda_{c}^{2}-2)(\sqrt{7}e^{i\delta_{cp}}-4), \nonumber\\
U_{\mu 1}&=& \frac{3}{32}(\lambda_{c}^{2}-2)\lbrace \sqrt{7}(1+\lambda_{c})-4\lambda_{c} e ^{-i\delta_{cp}}\rbrace, \nonumber
\end{eqnarray}
\begin{eqnarray}
U_{\mu 2}&=& \frac{1}{32}(2-\lambda_{c}^{2})( 9- 7\lambda_{c}+4\sqrt{7} e^{-i\delta_{cp}}), \nonumber\\
U_{\mu 3}&=& \frac{\sqrt{7}}{16}(\lambda_{c}^{2}-2)^{2} + \lambda_{c}^{2} e^{-i \delta_{cp}}, \nonumber\\
U_{\tau 1}&=&\frac{1}{16}(7-9\lambda_{c}),\nonumber \\
U_{\tau 2}&=& -\frac{3\sqrt{7}}{16}(1+\lambda_{c}),\nonumber\\
U_{\tau 3}&=& \frac{3}{8}(2-\lambda_{c}^{2}).\nonumber
\end{eqnarray}
Following Eq. (7) and from Eq. (13), we get
\begin{small}
\begin{eqnarray}
\sin^{2}\theta_{13} &=& \frac{1}{64}\lambda_{c}^{2}(\lambda_{c}^{2}-2)^{2}(23-8\sqrt{7}\cos\delta_{cp}),\\
\sin^{2}\theta_{12} &=& \frac{112+\lambda_{c}^{2}\lbrace7 \lambda_{c}(31\lambda_{c}-18)-143\rbrace+2\sqrt{7}\lambda_{c}(7\lambda_{c}-9)(\lambda_{c}^{2}-2)^{2}\cos\delta_{cp}}{256\lbrace1+\frac{1}{64}\lambda_{c}^{2}(\lambda_{c}^{2}-2)^{2}(8\sqrt{7}\cos\delta_{cp}-23)\rbrace}
,\nonumber \\
\\
\sin^{2}\theta_{23} &=& \frac{112-\lambda_{c}^{2}\lbrace224-424\lambda_{c}^{2}-32\sqrt{7}(\lambda_{c}^{2}-2)^{2}\cos\delta_{cp}}{256\lbrace1+\frac{1}{64}\lambda_{c}^{2}(\lambda_{c}^{2}-2)^{2}(8\sqrt{7}\cos\delta_{cp}-23)\rbrace}.
\end{eqnarray}
\end{small}
In the Ref [20], three data of $1\sigma$ ranges are specified regarding $\sin^{2}\theta_{23}$. They are 0.400-0.461 and 0.573-0.635 (N.H) and 0.569 - 0.626 (I.H). From Eq.(15), with the limit, $0\leq\vert \cos\delta_{cp}\vert\leq 1$, we get the bound of $\sin^{2}\theta_{23}$ as $0.427-0.463$ and hence out of all three possible $1\sigma$ bounds of $\sin^{2}\theta_{23}$, two are strongly ruled out and our analysis is very well fitted with the first one [fig.4]. This supports the existence of $\theta_{23}$ to lie within the first octant. It is to be noted that the best fit [20] of $\sin^{2}\theta_{23}$ i.e, $0.427$ coincides with our analysis when $\delta_{cp}=0$. 

From Eqs.(13 - 14), this is clear that Dirac phase $\delta_{cp}$ affects the prediction of all the three mixing angles which is different from BM case where only $\theta_{12}$ is affected by $\delta_{cp}$ (Eq. (9)). It seems that the situation is now much more complicated than the BM case. If our initial choice for $U_{\nu}$ as strictly BL and $U_{l}$ as CKM types were appropriate, then on placing the best fit [20] results at least for two of the three parameters in any two out of the three Eqs. (14-16), the predictions of $\delta_{cp}$ from the respective equations must coincide. The situation is as if for one unknown parameter $\delta_{cp}$, there are more than one equations. We first solve Eq.(14) with the best fit value of $\sin^{2}\theta_{13}$ [20], to find out $\cos\delta_{cp}$ and do the same for Eq.(15) with $ \sin^{2}\theta_{12}$. But surprisingly, we find the predictions of $\cos\delta_{cp}\approx 0.70$ (i.e $ \delta_{cp} \approx 0.25 \pi$) is same from both of the equations. In the next step, we put $\cos\delta_{cp}\approx 0.70$ in Eq. (16), and get $\sin^{2}\theta_{23}\approx 0.44$ which is close to best-fit result $\sin^{2}\theta_{23} =0.427$ [20]. These analyses are illustrated graphically in the figs.2-4. 

With strict BL mixing as the 1st approximation ( $U_{\nu}= U_{BL}(\lambda_{c}) $ ) and along with a unitary charged lepton mixing matrix ( $U_{l}(\lambda_{c},\delta_{cp})$ ) of CKM type, the predictions are summarised as follows.\begin{equation}
\sin^{2}\theta_{13}=0.0245,\quad \sin^{2}\theta_{12}=0.3209, \quad \sin^{2}\theta_{23}=0.4533, \quad \delta_{cp}= 0.2515\pi.
\end{equation}

From Eq. (13), we work out the CP violation Jarsklog invariant parameter as $J_{cp}^{BL}=Im[U^{*}_{e1} U^{*}_{\mu 1} U_{e3} U_{\mu 1} ]$,
\begin{align}
\vert J_{CP}^{BL} \vert = \frac{9\sqrt{7}}{4096}\lambda_{c}\lbrace 28-8\lambda_{c} (1+8\lambda_{c}-\lambda_{c}^{2})+ 57 \lambda_{c}^{4}\rbrace \sin\delta_{cp} \approx 0.0304 \sin\delta_{cp}.
\end{align}
If we choose $\delta_{cp}\approx 0.2515 \pi$, as per as the prediction, then we get $J_{CP}^{BL} \approx 0.0216$. 
\begin{figure}[t]
\begin{center}
\includegraphics[scale=1]{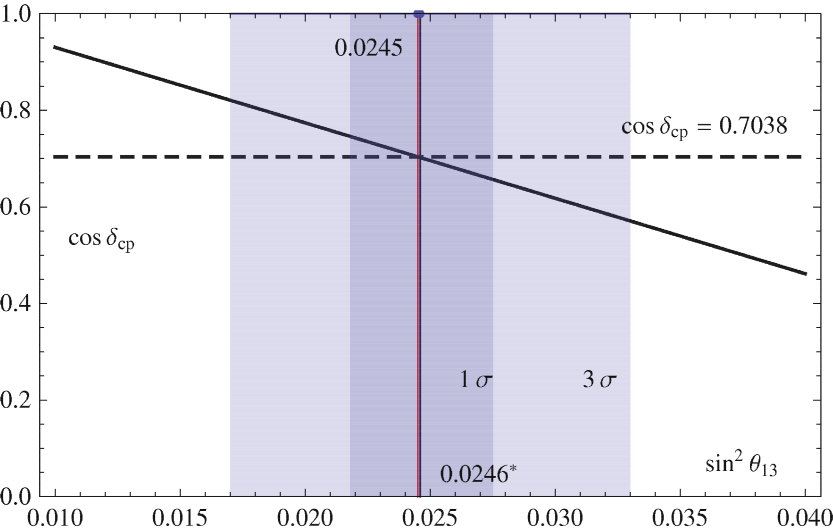} 
\caption{\footnotesize The dependence of $\cos\delta_{cp}$ on $\sin^{2}\theta_{13}$ for BL case with charged lepton correction.}
\end{center}
\end{figure}
\begin{figure}
\begin{center}
\includegraphics[scale=1]{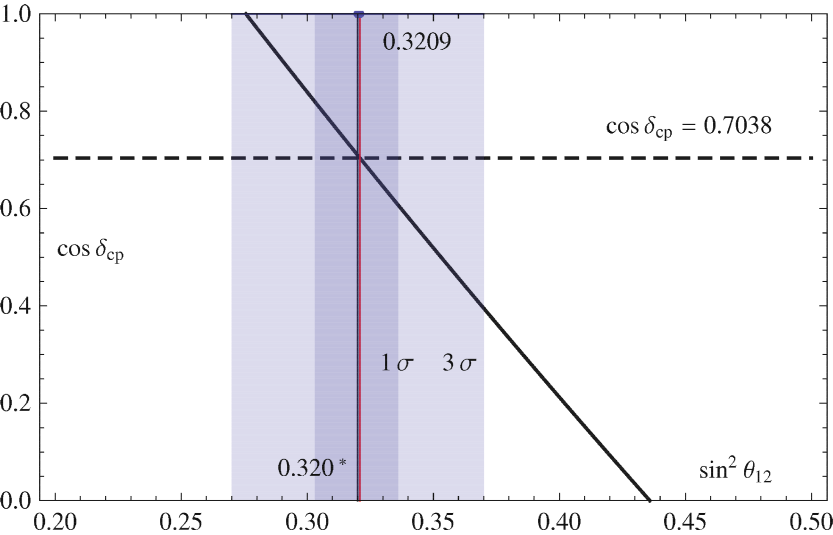} 
\caption{\footnotesize The dependence of $\cos\delta_{cp}$ on $\sin^{2}\theta_{12}$ for BL case with charged lepton correction.}
\end{center}
\end{figure}
\begin{figure}[t]
\begin{center}
\includegraphics[scale=1]{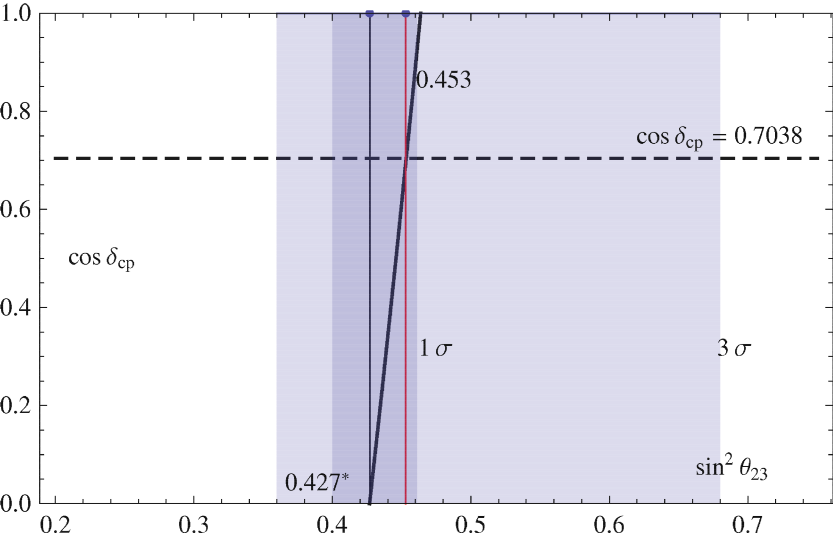} 
\caption{\footnotesize The dependence of $\cos\delta_{cp}$ on $\sin^{2}\theta_{23}$ for BL case with charged lepton correction.}
\end{center}
\end{figure}
\begin{figure}
\begin{center}
\includegraphics[scale=1]{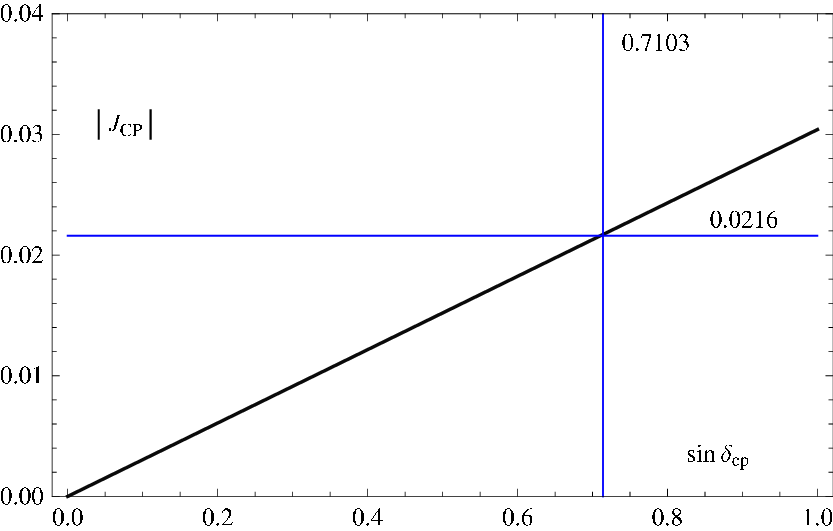} 
\caption{\footnotesize The variation of $J_{cp}$ with $\sin\delta_{cp}$ for BL case with charged lepton correction.}
\end{center}
\end{figure}
\section{Prediction of effective electron neutrino mass \textbf{$m_{ee}$ }in \textbf{$0\nu\beta\beta$} decay.}
The effective electron neutrino mass $m_{ee}$ appeared in neutrinoless double decay ($0\nu\beta\beta$)is given as
\begin{eqnarray}
m_{ee} = \vert m_{1} U_{e1}^{2} + m_{2} U_{e2}^{2} + m_{3} U_{e3}^{2} \vert
\end{eqnarray}
where $m_{i}$'s are the masses of the three neutrino mass eigenstates. Using Eq.(12), with $\lambda_{c}=0.2257$, and $\delta_{cp}\approx 0.2515 \pi$ we get
\begin{align}
m_{ee}=\vert 0.5262 m_{1} + 0.4056 m_{2} + 
 0.06954 m_{3} + (0.1953 m_{1} - 0.1314 m_{2} -0.0640 m_{3})\cos\delta_{cp} \vert.
\end{align}
For N.H case with $m_{1}$ as the smallest mass, we have,
\begin{eqnarray}
m_{2}=\sqrt{m_{1}^{2}+\Delta m_{21}^{2}},\quad m_{3}=\sqrt{m_{1}^{2}+\Delta m_{31}^{2}} .
\end{eqnarray}
We impose the Cosmological upper bound for $\Sigma m_{i} \leq 0.28 eV$ [21] in our analysis. We fix $\Delta m_{21}^{2} \sim 7.62 \times 10^{-5} eV^{2}$ (best-fit) [20] and $\Delta m_{31}^{2} \sim 2.55 \times 10^{-3} eV^{2}$ (best-fit)[20] and plot $\Sigma m_{i}$ taking lowest mass $m_{1}$ as free parameter and get the quasidegenerate upper limit for $m_{1}$ as 0.088 eV (fig. 6).
\begin{figure}[h]
\begin{center}
\includegraphics[scale=1]{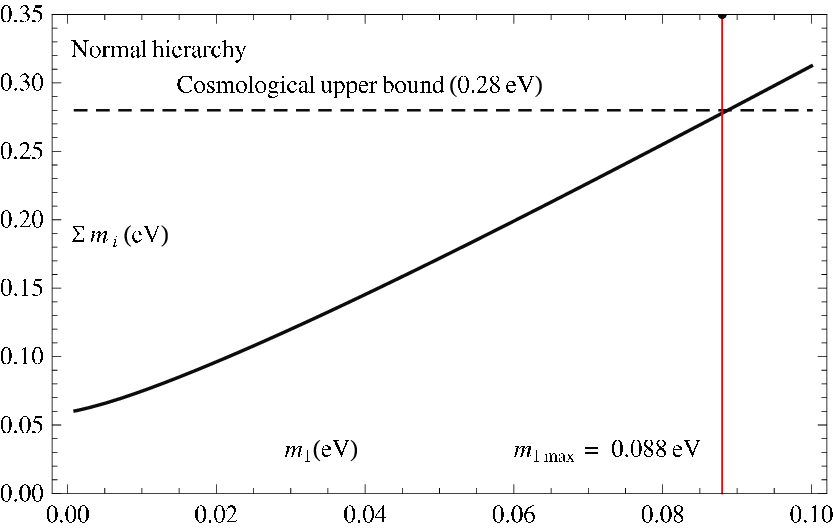}
\caption{\footnotesize The variation of $\Sigma m_{i}$ with $m_{1}$. The cosmological upper bound : $\Sigma m_{i}\leq 0.28 eV$. The Q.D limit of $m_{1}$ is $0.088 eV$.}
\end{center}
\end{figure}
We then plot $m_{ee}$ with respect to $m_{1}$ for three different cases concerning Majorana phases : ($ + m_{2}, + m_{3}$),( $ - m_{2}, + m_{3}$) and ( $ + m_{2}, - m_{3}$ ) (fig.7). Concerning this three cases the predictions for $m_{ee}$ under the quasidegenerate limit of $m_{1} \sim 0.088 eV$ are as follows.
\begin{eqnarray}
&( +m_{2}, +m _{3} )& : \quad \quad 0.0045 eV \leq m_{ee}\leq 0.0891 eV, \nonumber\\
&( -m_{2}, +m _{3} )&: \quad \quad 0\leq m_{ee}\leq 0.0335 eV,\nonumber\\
&( +m_{2}, -m _{3} )& : \quad \quad  0.0023 eV \leq m_{ee}\leq 0.0839 eV, 
\end{eqnarray}
where $\pm$ signs before $m_{2,3}$ indicate the Majorana CP phases.
\begin{figure}
\begin{center}
\includegraphics[scale=1]{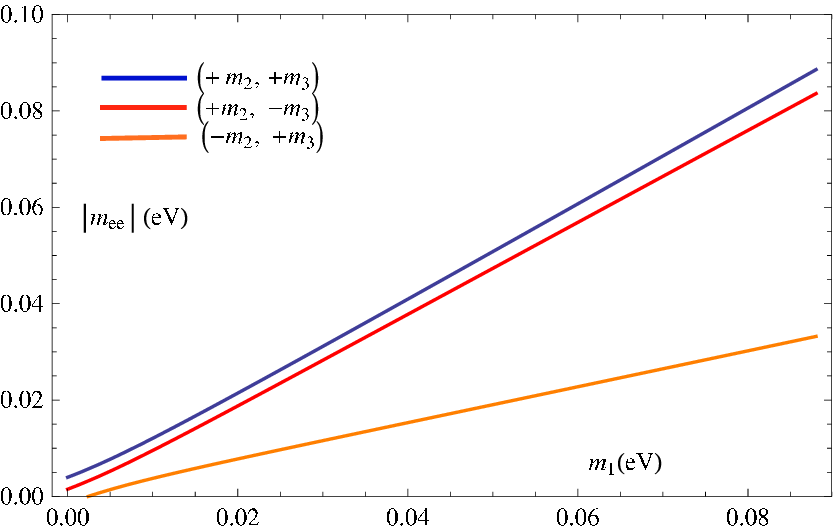}
\caption{\footnotesize The variation of $ m_{ee}$ with $m_{1}$ for ($+m_{2},+m_{3}$),$(-m_{2}, +m_{3})$ and $(+m_{2}, -m_{3})$ $CP$ cases.}
\end{center}
\end{figure}
Pascoli and Petcov[22] showed that if the neutrino mass ordering were of normal type, then $\vert m_{ee}\vert$ would satisfy $0.001 eV\leq\vert m_{ee}\vert$ which is consistent with the cases discussed above except ($ -m_{2}, + m_{3}$ ). There is an upper bound of neutrino mass parameter $ m_{ee}\leq 0.27 eV $ [23] which appears in the neutrinoless double beta decay experiments. The upper bounds of $m_{ee}$ for the three cases under quasidegenerate limit of $m_{1}$ satisfy this condition.
\section{Summary}
We have discussed the shortcomings of BM model where after considering the charged lepton correction, we are unable to lower the solar angle below $\sin^{2}\theta_{12}=0.3407$ (i.e,$\tan^{2}\theta_{12}=0.52$), although the prediction of $\theta_{13}$ and $\theta_{23}$ comply with the experimental results. Boucenna $et.al$ has itroduced a new mixing pattern called Bi-Large mixing where Cabibbo angle ($\lambda_{c}$) seeds the whole parametrization [Eqs.(1) -(3)]. We assume  $U_{l}$ to be of CKM type [Eq.(5)] and construct $U_{PMNS}$. This new model although phenomenological, is characterized by the following significant features: \textbf{(a)} Any other possibilities than $\theta_{23}$ to lie within the first octant, are sharply ruled out,\textbf{(b)} The predictions $\sin^{2}\theta_{13} \sim 0.0245$ and $\sin^{2}\theta_{12} \sim 0.3209$ are in precise agreement with the experimental best-fit values. We obtain $\sin^{2}\theta_{23}\sim 0.453$ which is close to the best-fit value (within $1\sigma$ range), and \textbf{(c)} $\delta_{cp}\sim 0.2515 \pi$ and $\vert J_{CP}^{BL} \vert = 0.0304 \sin\delta_{cp}\sim 0.0216$ . The same $U_{l}$ (CKM type), when incorporated with $U_{BM}$ was partly successful in complying with the experimental results because there it imposes a condition of $\delta_{cp}\rightarrow 0$,(i.e there is no CP violation) in order to lower the solar angle. Whereas this shortcoming is removed very easily when we associate the same CKM type $U_{l}$ with strict BL scheme. Hence the BL mixing scheme is very significant in the light of present experimental results. The model is further strengthened by the fact that the predictions of $\theta_{13}$, $\theta_{12}$ and $\theta_{23}$ individually  depend upon $\delta_{cp}$, without any contradiction. All the three angles agree to the desired results for a single choice of $\delta_{cp}\sim 0.2515 \pi$. Finally the model is employed to study the upper  bounds of $m_{ee}$ in quasidegenerate limit for three different Majorana $CP$ phases of normal hierarchy. A formal derivation of BL mixing matrix from discrete symmetry is an important aspect for our future investigation.
\section*{Acknowledgement}
One of the authors (SR) wishes to convey his heartiest gratitude to Chandan Duarah of Department of Physics, Gauhati University for usefull discussion. 

\end{document}